\documentstyle[aps]{revtex}

\begin{document}

\draft
\title{Numerical studies of the extended two-chain model of friction}
\author{Mario Einax, Steffen Trimper, H.-R. H\"oche and Michael Schulz}
\address{Fachbereich Physik, Martin--Luther--Universit\"at\\
Halle,06099 Halle (Saale), Germany}
\date{\today}
\maketitle

\vspace*{7ex}
\noindent
\begin{abstract}
\noindent
We investigate numerically a simple microscopic model to 
describe wearless dry friction between atomically flat contact
interfaces without thermal fluctuations $(T=0\,K)$. The
analysis of the incommensurate ground state shows a breaking of
analyticity when the amplitude $\epsilon$ of the Lennard-Jones 
interaction between the both harmonic chains increased beyond a
critical value $\epsilon_c$. By the introduction of a suitable 
order parameter and using a finite size scaling we could show 
that the breaking of analyticity is a second order phase transition.
This transition is often called Aubry transition. By applying of an
uniform external force $F^{ex}$ we have determined the critical 
force of static friction $F_c$ above a sliding motion occurs. 
\end{abstract}


\section*{{\normalsize {\bf I. INTRODUCTION }}}
\noindent
There are several ways for studying friction phenomena. Experimental
observations are very powerful to characterize the frictional behaviour
and to formulate empirical friction laws, e.g., Coulomb-Amonton's laws
\cite{b1}, from a phenomenological point of view. Macroscopically, everybody
can usually observe that during two solids are slided against each other, wear
take place. But wear is not necessary for the occurrence of friction. In the
last few years the technological development makes it possible to study
wearless dry friction between atomically flat contact interfaces experimentally
(nanotribology) \cite{b1,b2,b3}. These nano-scale experiments on systems with
less complicity stimulated again the theoretical efforts to investigate
friction phenomena on the base of microscopic models
\cite{b4,b5,b6,b7,b8,b9,b10,b11,b12}. The dissipative nature of friction is
a typical non-equilibrium problem. From the theoretical
point of view, wearless dry friction between two atomic flat solids
could be explained by using a simple driven mechanical many-body system
with a lattice structure and take into account interaction forces.
Thus the question arises, how the structure of such a
microscopic model of friction has to look.

One of the simplest structure on the microscopic level is given by the
discrete version of the well-known Frenkel-Kontorova (FK) model \cite{b13}, 
which is in general a model of an one-dimensional harmonic chain of 
particles in a periodic potential. The FK model is mainly used for
describing an adsorbate atom monolayer on an atomically flat
clean substrate surface. Using this picture, the driven FK model, where
a constant external force is applied to each particle of the chain,
can also be used as a simple model structure for a wearless friction 
situation, which has been studied by several works \cite{b4,b5,b6}.
Note, several physical phenomena such as the dislocations motion in
crystals \cite{b14}, commensurate-incommensurate transitions \cite{b15,b16} 
and charge-density waves (CDW`s) \cite{b17,b18} could also be investigated
by reducing on the FK model structure. Definitely, the simple FK model 
is a good starting point in order to construct more complex and
realistic friction models on the atomic level. Thus, a natural extension 
of the FK model is the replacement of the rigid substrat (hard body) by a 
deformable substrat monolayer pinned on a bulk (soft body). Such a more 
complex model of friction was firstly introduced by Matsukawa and Fukuyama 
\cite{b7,b8}. They considered an one-dimensional model consisting of 
two deformable interacting harmonic chains, where each particle of the lower 
chain is harmonically pinned on a rigid bulk. The static and kinetic 
properties of this two-chain model of friction was analysed in great details 
by Kawaguchi and Matsukawa \cite{b9,b10}. So far, the FK model and the 
two-chain model are simple models of wearless friction between an atomic 
monolayer and a hard or soft body, respectively. For describing a more 
realistic friction situation, which should represent wearless friction 
between atomically flat bodies, it is necessary to replace the adsorbate 
monolayer by a substrat. For this reason, Weiss and Elmer \cite{b11,b12} 
proposed an interesting approach, the so-called Frenkel-Kontorova-Tomlinson 
(FKT) model, which is an one-dimensional lattice model for a soft upper body
sliding on a hard lower body. In contrast to the two-chain model, with the 
FKT model they extended the FK model structure in another direction by 
harmonically coupling of each particle of the adsorbat monolayer on an upper 
macroscopic sliding mass, whose position relative to the lower surface is 
characterized by the coordinate $x_B$. Note, all of these models are 
investigated as completely one-dimensional microscopic friction models, where 
the particles of the one-dimensional chain can only move parallel to the 
direction of the external force. Now, it is obviously clear, how you have to 
extend the previous model structures in order to write down a more complex 
microscopic model system, which is necessary to describe friction phenomena as 
simple but also realistic as possible. For that purpose we introduce an extended 
two-chain model consisting of two interacting harmonic atomic chains embedded 
in a two-dimensional space, where each particle of the upper or lower chain 
is harmonically pinned on an upper or a lower rigid bulk, respectively. 
This model describes two soft bodies with an one-dimensional contact 
interface, which can carry out a genuine two-dimensional sliding motion  
against each other. In our studies we consider the extended two chain model 
as a simple relaxation model by taking energy dissipation via a 
phenomenological damping term in the mechanical equation of motion into account. 
Hence, all model, where the energy dissipation is modeled by hand through an
explicit damping term,  cannot explain the origin of friction in complex 
physical systems refering to the sliding state. Well, that is also not the aim 
of our investigations, because with our approach we want to establish the 
typical macroscopic friction situation (tension experiment) on a microscopic
level. Nevertheless, the introduced model structure can be used to discuss
the mechanism of the occurrence of static friction, which is one characteristic
feature of dry friction, based on an energy concept. The hope is that the 
properties of the ground state - defined as the minimal potential energy state 
of the mechanical model system in absence of an external force - describe 
completely the static friction behaviour as function of the interaction 
between the two deformable chains. In general the ground state structure can 
exhibit two situations as a driven force increased from zero. On the one hand, 
a zero static friction arises exactly then if the maximum static friction force 
vanished, i.e., the model system is frictionless. On the other hand, 
a finite static frition corresponds always with the existence of a finite 
maximum static friction force, which is the maximum potential energy state 
of the system above which a sliding state appears. In other words, if we
evolve the particle configuration from their ground state configuration, then 
the maximum static friction force is defined by the largest depinning force, 
which is equivalent to the smallest driven force above which no stable 
stationary state exists. 

For the analysis of static properties of the extended two chain model we follow 
the philosophy of Aubry by using strangely enough an incommensurate ground 
state structure, which has the only purpose to introduce a special kind of 
mathematical functions called as hull-functions. On the basis of the specific
hull-function concept for the visualisation of the ground state structure of 
the discrete FK-model in the incommensurate case Aubry \cite{b19,b20} has been 
shown that a new type of phase transition (called as breaking of analyticity) 
occurs when the amplitude of the periodic potential increased beyond a 
critical value. That means, below a critical amplitude the hull-functions have 
a continuous (analytic) structure whereas above a critical amplitude the 
hull-functions are discrete (non analytic). In our numerical investigations of 
the hull-functions of the extended two chain model we have also found the 
transition by breaking of analyticity (Aubry transition). In order to 
distinguish, whether the Aubry transition is really about a second order phase 
transition, we have specifically studied this type of phase transition like 
Coppersmith and Fisher (but which have been investigate the behaviour near
the maximum static friction force \cite{b17,b18}) from the viewpoint of 
critical phenomena. For that purpose we have introduced a suitable order 
parameter, which shows the typical critical behaviour for second order phase 
transitions and critical exponents can be calculated via scaling arguments.

Here, it should mentioned, that from the physical intuition a finite maximum 
static friction as function of the interaction strength $\epsilon$ has to 
exist in all previous introduced models equal whether one choose a commensurate 
or an incommensurate ground state structure. The main differences are the 
following. In the commensurate case the static friction is always larger than 
zero for any $\epsilon > 0$. By way of contrast, in the incommensurate case
whether the static friction is equal or unequal zero depends strongly on 
$\epsilon$ below or above a critical value $\epsilon_c$ due to the preculiar
breaking of analyticity. Hence the commensurate and the incommensurate ground 
state structure reveals two types of physical system classes characterizing 
by different critical behaviour in the sliding state near the threshold 
force $F_c$. The first class is given by model systems with a commensurate
ground state structurce. Effectively, commensurate models are characterized by 
finitely many internal degrees of freedom, which lead to trivial critical
exponents in the sliding state for the velocity-force characteristic near the 
threshold $F_c$. An incommensurate ground state structure is equivalent to a 
second class of models with infintely many internal degrees of freedom. 
Well, both model classes emphasize the fact that the nature of static and 
kinetic friction is deeply rooted with internal degrees of freedom generated 
by a complex atomic interaction behaviour, which is the key for unterstanding 
dissipation in general. Commensurate model have one  main disadvantage, because
an uniquely visualisation concept for the ground state structure like the 
hull-function for the incommensurate case is missing. Moreover,
commensurate models exist always in a pinned ground state structure, that means,
models imply static friction for any value of the amplitude of the interaction 
strength larger then zero intrinsicly. Therefore, for understanding the 
physical fundamental mechanism of the arising of static friction it seems to 
be most interesting to study incommensurate models, because the ground state 
shows two different incommensurate phases related to amazing properties. That 
means, an unpinned incommensurate phase corresponds to zero static friction 
whereas the pinned incommensurate phase is equivalent to a finite static 
friction.

In the present paper we introduce the extended two chain model and the 
numerical calculation methods. The detailed investigations of the static 
properties, especially the ground state  structure of the extended two chain 
model are discussed by defining an order parameter and proposing scaling
relations.

\section*{{\normalsize {\bf II. THE MODEL SYSTEM}}}
\noindent
Following Matsukawa and Fukuyama \cite{b7,b8}, we introduce the extended
two-chain model, which is a two-dimensional microscopic lattice model system 
describing wearless dry friction between two atomically flat solids. Therefore, 
we consider two deformable atomic chains, so called an upper and a lower chain.
The atoms in each chain are classical point particles and are able to realize
a two-dimensional movement. We assumed a harmonic particle interaction in each 
chain. The interaction between the two chains is of the Lennard-Jones type. 
Moreover, both chains are coupled harmonically on an upper or a lower substrat
respectively. Like Ref. \cite{b7} the effects of energy  dissipation in the 
upper and lower chain are phenomenologically approached to be proportional to 
the difference between the velocity of the $i$-th particle and that of the 
center of gravity. The external force ${\underline F^{ex}} = (F^{ex}, 0)$ acts on the 
upper chain in $x$ direction without $y$ component. In order to simplify the 
situation, we consider overdamped motion. The equations of motion for the 
particles in the upper and the lower chain are given for each direction as 
follows (see also FIG. 1)
\begin{eqnarray}
\label{xakette}
m_a \gamma_a \left( \, {\dot x}_i^a - \left< {\dot x}_i^a \right> \, \right)
&=& k_a^x \big[  \, x_{i-1}^a + x_{i+1}^a - 2 x_i^a \big] - 
k_{sub}^{a_x} \big[ x_i^a - i c_a \big] +
\sum_{j=1}^{N_b} F_{int}^{x} \left( x_i^a - x_j^b \right) +F^{ex} \\
\label{yakette}
m_a \gamma_a \left( \, {\dot y}_i^a - \left< {\dot y}_i^a \right> \, \right)
&=& k_a^y \big[  \, y_{i-1}^a + y_{i+1}^a - 2 y_i^a \big] - 
k_{sub}^{a_y} \big[ y_i^a - (b_b+L_0) \big] +
\sum_{j=1}^{N_b} F_{int}^{y} \left( y_i^a - y_j^b \right) \\
 & & \nonumber \\
\label{xbkette}
m_b \gamma_b \left( \, {\dot x}_i^b - \left< {\dot x}_i^b \right> \, \right)
&=& k_b^x \big[  \, x_{i-1}^b + x_{i+1}^b - 2 x_i^b \big] - 
k_{sub}^{b_x} \big[ x_i^b - i c_b \big] +
\sum_{j=1}^{N_a} F_{int}^{x} \left( x_i^b - x_j^a \right) \\
\label{ybkette}
m_b \gamma_b \left( \, {\dot y}_i^b - \left< {\dot y}_i^b \right> \, \right)
&=& k_b^y \big[  \, y_{i-1}^b + y_{i+1}^b - 2 y_i^b \big] - 
k_{sub}^{b_y} \big[ y_i^b - b_b \big] +
\sum_{j=1}^{N_a} F_{int}^{y} \left( y_i^b - y_j^a \right)  \,  , 
\end{eqnarray}
where $\underline r_i^a=(x_i^a, y_i^a)$ and $\underline r_i^b=(x_i^b, y_i^b)$ are the immediately
position vectors of the $i$-th particle in the upper and lower chain. 
Furthermore, $N_a$ ($N_b$), $m_a$ ($m_b$) is the particle number and the 
particle mass, $\gamma_a$ ($\gamma_b$) denotes the friction coefficient
and $c_a$ ($c_b$) is the mean atomic spacing in $x$-direction of the
upper (lower) chain. It should be mentioned, that the particle mass 
and the friction coefficient are for all particles in each chain the 
same. For a compact representation one can introduce the chain index 
$I=(a,b)$ relating to the upper and lower chain. The velocity of the 
center of gravity is given by the average 
$\left< {\dot X}_i^I \right> = \frac{1}{N_I} \sum_{i=1}^{N_I} {\dot X}_i^I$, 
where ${\dot X}_i^I$ stands for ${\dot x}_i^I$ and ${\dot y}_i^I$
respectively. $k_I^x$ and $k_I^y$ are the strength of the harmonic
interaction force in the chains for the $x$ and $y$ direction. The strenght 
of the coupling of each chain on an upper and a lower substrat
is represented by $k_{sub}^{I_x}$ and $k_{sub}^{I_y}$, which
determine the rigiditiy of the chains. A further system parameter is
the substrat-substrat distance $L = b_a + b_b + L_0$, where $b_a$
($b_b$) is the mean atomic spacings in $y$-direction between the 
upper (lower) chain and upper (lower) substrat. $L_0$ denotes the 
distance between the two harmonical chains and is a function of
their interaction strength. The interaction force between the upper 
and lower chain, projected on the $x$ and $y$ axes, can be written as
\begin{eqnarray}
\label{xww}
\sum_{j=1}^{N_J} F_{int}^{x} \left( x_i^I - x_j^J \right) &=&
\sum_{j=1}^{N_J} F(r_{ij}) \, \frac{\left( x_i^I - x_j^J \right)}{r_{ij}} \\
 & & \nonumber \\
\label{yww}
\sum_{j=1}^{N_J} F_{int}^{y} \left( y_i^I - y_j^J \right) &=&
\sum_{j=1}^{N_J} F(r_{ij}) \, \frac{\left( y_i^I - y_j^J \right)}{r_{ij}} \,
\end{eqnarray}
with the chain indices $I=(a,b)$, $J=(b,a)$ and the distance
$r_{ij}=\left| {\underline r}_i^a - {\underline r}_j^b \right| \equiv \sqrt{( x_i^b - x_j^a )^2 + ( y_i^b - y_j^a )^2}$ relating
to the $i$-th atom of the upper (a) and the $j$-th atom of the lower (b)
chain. $F(r_{ij})$ can be expressed by the potential $V(r_{ij})$ through
$F(r_{ij})=-\frac{\partial V(r_{ij})}{\partial r_{ij}}$. As interaction potential 
we choose specifically the Lennard-Jones potential
\begin{equation}
V(r_{ij}) = 4 \epsilon \left[ \left(\frac{\sigma}{r_{ij}} \right)^{12} - 
\left( \frac{\sigma}{r_{ij}} \right)^6  \right] 
\end{equation}
representing the character of the friction due to the intermolecular 
forces in the best way. Here $\epsilon$ is the strength of the interaction 
and $\sigma$ is a characteristic length parameter, which parameterized 
the minimum of the potential 
$V^{min} = \left. V(r_{ij}) \right|_{r_{ij} = r_0}$
via the distance from equilibrium $r_0= \sigma \, 2^{1/6}$.

\subsection*{{\normalsize {\bf III. NUMERICAL METHOD}}}
\noindent
For solving the two-dimensional extended two-chain lattice 
model we choose periodic boundary conditions in both chains
\begin{eqnarray}
\label{rb}
 x_{i}^{I} &=& x_{i+N_I}^{I} - L_I  \\
 y_{i}^{I} &=& y_{i+N_I}^{I}
\end{eqnarray} 
with the chain index $I=a,b$ and the chain lengths $L_I=N_I c_I$.
This implies that the system size of the upper and lower chain are
the same $L \equiv N_a c_a = N_b c_b$. Hence, the ratio or the
misfit of the two lattice constants could be expressed by the 
particle numbers
\begin{equation}
\alpha = \frac{c_a}{c_b} = \frac{N_b}{N_a} \, .
\end{equation}
This means, that the ratio $\alpha$ of the two lattice constants
is always a rational number in the numerical realisation. Using the 
concept of the hull-function, which was introduced by Aubry and
coworkers \cite{b19,b20} for describing the ground state of the 
underlying model system, we need an irrational ratio $\alpha$
between $c_a$ and $c_b$. If we have a rational $\alpha$, i.e., in the
most commensurable case $\alpha =1$, we cannot apply the concept of
the hull-function. Because the whole ground state configuration
of each chain, consisted of $N_I$ particle positions, would be 
represented by one value of the argument of the hull function and
hence the hull-function plot lose his mathematical power of proposition.
For this reason, it is necessary to investigate an incommensurate
ground state structure relating to the ratio $\alpha$ of the two 
lattice constants. Unfortunately, in the numerical calculation we 
could only approximate an irrational number through optimal 
rational ratios of two natural numbers using a finite 
continued-fraction expansion up to a certain order \cite{b18,b19}
\begin{equation}
\alpha = a_0 + \frac{1}{\displaystyle a_1
              + \frac{1}{\displaystyle a_2
               + \frac{1}{\displaystyle \ldots
		+ \frac{1}{\displaystyle a_{n-1}
		 + \frac{1}{\displaystyle a_n}}}}}
\end{equation}
Following the literature, we choose the simplest continued-fraction
expansion with $a_0=1$ and $a_1=a_2= \cdots = a_n =1$, so that 
$\alpha$ could be expressed by the ratios of the Fibonacci numbers 
$F_{n+1} = F_n + F_{n-1}$, where $n=1,2,3,\cdots$ as well as $F_0=1$ 
and $F_1=1$. Hence,
\begin{equation}
\label{fibonacci}
\alpha = \frac{F_{n+2}}{F_{n+1}} = \frac{3}{2} \, , \frac{5}{3} \, ,
\frac{8}{5} \, , \frac{13}{8} \, , \frac{21}{13} \, , \frac{34}{21} \, ,
\frac{55}{34} \, , \frac{89}{55} \, , \frac{144}{89} \, ,
\frac{233}{144} \, , \frac{377}{233} \, , \frac{610}{377} \, ,
\frac{987}{610} \, , \frac{1597}{987} \, , \, \cdots
\end{equation}
and the irrational limit $\alpha \equiv \lim_{n \to \infty} 
\frac{F_{n+2}}{F_{n+1}} = \frac{1}{2} \left( \sqrt{5} + 1 \right)$ of 
the approximation series (\ref{fibonacci}) is the inverse of the so-called 
{\it golden mean}. From the point of number theory \cite{b19} , 
this is the most incommensurable case of the underlying model system,
which could be emulated in a systematical way. In our studies we set
the lattice constant of the lower chain equal one ($c_b=1$).
Accordingly, the upper mean lattice spacing $c_a$ is equal
$\alpha$ and we calculated with the system sizes
\begin{equation}
\label{systemsize}
c_a = \frac{N_b}{N_a} = \frac{89}{55}\, , \frac{233}{144} \, ,
\frac{377}{233} \, , \frac{610}{377} \, , \frac{1597}{987} \, .
\end{equation}
It should be mentioned, that not all ratios oft the approximants
are "good" irrational number approximations in the same manner
because for certain particle number ratios there exist some
artifacts in the hull function. \\
\noindent
For the numerical calculation of the equations of motion we used 
a fourth-order Runge-Kutta algorithm with an adaptive stepsize 
control. In order to discuss the essential features of the present 
model we restrict the wide range of model parameters by setting 
$m_I = 1$, $\gamma_I = 1$, $k_I^x = k_I^y = 1$, 
$k_{sub}^{I_x} = k_{sub}^{I_y} = 1$ and $\sigma = 1$. 
As initial conditions we assumed at $t=0$ a particle configuration 
in the x-y-plane, where the $N_a$ and $N_b$ chain particles are
arranged at regular lattice sites periodically. Besides, the
distance $L_0$ is set to an arbitrary initial value, i.e. $L_0=1$. 
For calculating the stable stationary states we used the same
criterion like Kawaguchi and Matsukawa \cite{b10}. That means, 
we used a velocitiy condition 
\begin{equation}
\label{cond}
\sqrt{\frac{\left( \sum_{i=1}^{N_a} (\dot{r}_i^a)^2 +
\sum_{j=1}^{N_b} (\dot{r}_j^b)^2 \right)}{(N_a+N_b)}} \, < \, 10^{-5}
\end{equation}
for stopping the Runge-Kutta calculation. Hence, the system could be 
considered to be in a static state. 
The next step consists in the check up of the total 
sum of the coupling forces in the (a)-chain and (b)-chain in 
$x$-direction
\begin{equation}
\label{pinnx}
F_{sub}^{a_{x}} = \sum_{i=1}^{N_a} \, k_{sub}^{a_x} 
\big[ \, x_i^a - i c_a \, \big] ~~~ , ~~~ 
F_{sub}^{b_{x}} = \sum_{j=1}^{N_b} \, k_{sub}^{b_x} 
\big[ \, x_j^b - j c_b \, \big]
\end{equation}
and $y$-direction
\begin{equation}
\label{pinny}
F_{sub}^{a_{y}} = \sum_{i=1}^{N_a} \, k_{sub}^{a_y} 
\big[ \, y_i^a - (b_b + L_0) \, \big] ~~~ , ~~~ 
F_{sub}^{b_{y}} = \sum_{j=1}^{N_b} \, k_{sub}^{b_y} 
\big[ \, y_j^b - b_b \, \big]
\end{equation} 
respectively. In absence of a driven force ($F^{ex}=0$) it is 
sufficient to consider only the value of the total sum of the 
coupling forces (\ref{pinny}) in $y$-direction, because for $t=0$,
after the Runge-Kutta steps and in the final state the
$x$-direction forces (\ref{pinnx}) are equal to zero. The 
regular periodical arrangement of the particles for the initial
time $t=0$ leads to $F_{sub}^{a_{y}}=F_{sub}^{b_{y}}=0$. After the
relaxation in the final stationary state relating to the velocity
condition (\ref{cond}) the total sum of coupling forces (\ref{pinny})
must always be zero, which is in general not true for an arbitrary
chosen  initial value $L_0$. Accordingly, during the numerical
calculation every final stable state in respect to $L_0$ must be
proofed and if necessary $L_0$ must be changed like
$\L_0 = L_0 \pm \delta L_0$. Then the chains relax again and the
procedure would be repeated  until the system reached a final 
stationary state, which fulfil the condititon 
$F_{sub}^{a_{y}}=F_{sub}^{b_{y}} < 5 \cdot 10^{-5} ~ ( \, \simeq 0 \, )$.
In this way we could find numerically the stationary configuration
of the extended two-chain model and for $F^{ex}=0$ one can assumed
that this is approximately the ground state, defined as having an
energy which cannot be lowered by moving of some system particles.
The properties of the ground state could be analysed by changing
the interaction strength $\epsilon$, which corresponds to set the
correct $L_0=L_0 (\epsilon)$ numerically. \\
In the presence of an external force $F^{ex} > 0$ acting in
$x$-direction one must slightly change the numerical calculation
procedure. Instead of $F_{sub}^{a_{x}}=F_{sub}^{b_{x}}=0$
the $x$-direction forces (\ref{pinnx}) have for $t=0$ the values
$F_{sub}^{a_{x}}=F^{ex}$ and $F_{sub}^{b_{x}}=0$ respectively. 
After each Runge-Kutta step we have to guarantee that the control 
parameters (\ref{pinnx}) have again these values. Hence, we are able
to evolve the model system from the ground state into the last 
possible static solution by continuously increasing of the applied 
external force. In this sense we could numerically define the threshold, where no longer a static solution of the underlying system
exist. This threshold $F^{ex}=F_c$ is the critical force of static 
friction, which is also called maximum static frictional force. Below 
$F^{ex}< F_c$ the chains are in a steady state, such that the afterwards
space-time averaged velocity $v \equiv \left< {\dot x}_i^a\right>_{i,t} = 0$.
Above $F^{ex}>F_c$ the upper chain, which is after an initial transient decay
in a steady sliding state, slides with a constant velocity $v$. Note, in an easy
way one can investigate the present model in a gravitational field via 
the constraints (\ref{pinny}).

\section*{{\normalsize {\bf IV. NUMERICAL RESULTS}}}

\subsection*{{\normalsize {\bf A. Ground state}}}
\noindent
First, we investigate the stationary states of the extended two-chain model in 
the absence of an external force, i.e., $F^{ex}=0$. The stationary states are
given by the solution of equation system (\ref{xakette}-\ref{ybkette})
with vanishing right-hand side 
\begin{equation}
\label{stationary}
0=  m_a \gamma_a \left( \, {\dot {\underline r}}_i^a - \left< {\dot {\underline r}}_i^a \right>
\, \right) = m_b \gamma_b \left( \, {\dot {\underline r}}_i^b - \left< {\dot {\underline r}}_i^b 
\right> \, \right)
\end{equation}
with consideration of the conditions (\ref{pinnx}) and (\ref{pinny}), where 
$F_{sub}^{a_{x}}=F_{sub}^{a_{y}}=F_{sub}^{b_{x}}=F_{sub}^{b_{x}}=0$.  
Hence, the ground state of the extended two chain model is the stationary
particle configuration having the minimal potential energy for $F^{ex}=0$.
The knowledge of the ground state determine a physical system, such as the 
static friction, completely.

\subsubsection*{{\normalsize {\bf 1. The hull-function}}}
\noindent
Similarly to the FK model \cite{b19} or FKT model \cite{b11}, the lattice
structure of the particle positions in the incommensurate ground state can be 
uniformly described by a hull-function. For examining the lattice structure of 
the present two-dimensional extended two-chain model, it is necessary  to 
introduce four hull-function due to two degrees of freedom of the particles in 
each chain. The two hull-functions $h_a$ and $h_c$ of the upper chain are 
defined as
\begin{eqnarray}
\label{ha}
x_i^a &=& i \, c_a + \phi + h_a ( i \, c_a +\phi) \\
\label{hc}
y_i^a &=&  b_b + L_0 + \phi ' + h_c ( i \, c_a + \phi ') 
\end{eqnarray}
and the two hull-function $h_b$ and $h_c$ of the lower chain are given by
\begin{eqnarray}
\label{hb}
x_i^b &=& i \, c_b + \psi + h_b ( i \, c_b + \psi) \\
\label{hd}
y_i^b &=& b_b + \psi ' + h_d ( i \, c_b + \psi ')  \, \, \, ,
\end{eqnarray}
where $\phi$, $\phi '$, $\psi$ and $\psi '$ are constant phases. For our 
investigations, it is conviencent to set $\phi = \phi ' = \psi = \psi ' =0$. 
The hull-functions are periodic and even, which can usually be expressed by
\begin{eqnarray}
\label{pupper}
h_a (z + c_b) &=& h_a (z) = - h_a (-z), \; \; \; \;  h_c (z + c_b) = h_c (z) = 
- h_c (-z) \\
\label{plower}
h_b (z + c_a) &=& h_b (z) = - h_b (-z), \; \; \; \;  h_d (z + c_a) = h_d (z) = 
- h_d (-z) \, .
\end{eqnarray}
The argument $z$ of $h_a$ and $h_c$ is defined in the range from $0$ to $c_b$,
whereas of $h_c$ and $h_d$ it is given from $0$ to $c_a$. Note, the 
visualisation concept of the hull-function makes it necessary that the argument
of the just defined hull-functions $h_a$, $h_c$ or $h_b$, $h_d$ must be the 
same, respectively. Because, the displacements in $x$- resp. $y$-direction of 
all particle in each chain can be assigned to different values of the argument 
$z$ due to the irrational ratio $\alpha$ of the mean lattice spacing $c_a$ and 
$c_b$. 
 
First, we have numerically calculated the ground state of the extended 
two-chain model for a fixed ratio $\alpha=N_b/N_a $ of the particle numbers 
given by $N_a =233$ and $N_b=377$. Several investigations of the FK model 
\cite{b20}, FKT model \cite{b11} such as the two-chain model \cite{b10} showed, 
that the ground state strongly depends on the interaction strength $\epsilon$. 
Therefore, we started the computation for different values of $\epsilon$ in the 
wide range from $0.05$ to $0.4$. For small values of $\epsilon$ up to a certain
value $\epsilon_c$ we get the result that the periodic hull-functions $h_a$, 
$h_b$, $h_c$ and $h_d$ are analytic, i.e, smooth and continuous. This behaviour 
is illustrated in Fig. 2, where the interaction strength has the value 
$\epsilon=0.22$. Furthermore, the stationary state for $\epsilon=0.22$ is
characterized by the distance $L_0=0.986273$ between the upper and lower chain. 
Figure 3 shows another numerical calculation of the hull-functions for the
interaction strength $\epsilon=0.33$ with the corresponding distance 
$L_0=0.959584$. We can see that the hull-functions in Fig. 3 are discrete, which 
means that they are no longer analytic. This discrete structure is related to 
jumps, which occur at certain points of the argument $z$ of the hull-functions.
Now, the comparison of Fig. 2 with Fig. 3 shows the phenomenon of breaking of
analyticity, which is due to two significant aspects of the hull-functions 
$h_a$, $h_b$, $h_c$ and $h_d$ for the interaction strength $\epsilon < \epsilon_c$ 
and $\epsilon > \epsilon_c$, which defined a certain transition value 
$\epsilon_c$ above the hull-functions are no longer analytic. For a detailed 
analysis of the threshold $\epsilon_c$ we introduced a suitable order parameter
in the following part. 

Well, Aubry has shown \cite{b20} that the breaking of analyticity of the
incommensurate ground state of the FK model is characterized by the existence
of a largest central gap at the half of the period of the hull-function. Whereas
the situation in the two-chain model of Matsukawa and Fukuyama \cite{b10} is
quite different for several elastic parameters and three simple cases are 
observed: both hull-functions $h_a$ and $h_b$ have a largest gap at the half of 
the period (FK limes), whether $h_a$ or $h_b$ has only a largest central gap 
and finally both hull-functions do not have a largest central gap but two 
symmetrical gaps with the same size refering to the half of the period. 
In Fig. 3 we can also observe that no largest central gap for all hull-functions
of the extended two-chain model exists. Furthermore, the hull-functions show 
also two symmetrical gaps of the same size refering to the half of the period 
of the hull-function in the whole analysed parameter range. Up to now we have 
only investigated the ground state structure for a fixed system size 
($N_a=233$ and $N_b=377$) and a certain range of interaction strength 
$\epsilon$. Now, we vary the system size of the underlying model which
corresponds the different approximation orders of the irrational number through
a rational ratio of the particle numbers. Here, we calculated especially the 
ground state structure for the ratios $\alpha=89/55$, $233/144$, $610/377$, 
and $1597/987$ in the same range of the interaction strength $\epsilon$ like
above. In particular, we proofed whether there occurs an anormalous behaviour in
the structure of the hull-functions for certain values of $\epsilon$ due to  
the using of a rational ratio instead of an irrational number in the numerical
computation. Indeed, Fig. 4 shows such a anormalous behaviour of the
hull-functions for $\epsilon=0.33$ and for the particle numbers $N_a=377$ and
$610$. In the comparison with Fig. 3 we can recognize that the even particle 
number $N_b=610$ generates a wrong particle position at the half of the 
period $x=0.5$ of $h_b$ in Fig. 4 due to the periodical boundary condition.
Consequently, in term of a suitable rational particle ratio Fig.\,3 shows the
right picture of the hull-functions for $\epsilon=0.33$, whereas Fig. 4 shows
some artifacts in the hull-function structure for the same interaction strength
$\epsilon=0.33$ as a result of bad choice $N_a=377$ and $N_b=610$ of the 
system size for the interaction strength $\epsilon=0.33$. Beyond it, there
appear much more artifacts in the hull-function structure different ranges 
of $\epsilon$ als consequence of a bad choice for the particle number ratios.
For $N_a=144$ and $N_b=233$ the occurence of such artifacts destroyed the order
parameter analysis of the transition area. Kawaguchi and Matsukawa interpreted 
in \cite{b10} these artifacts, which also occur in the two-chain model for 
certain values of the interchain strength and $N_a=144$, $N_b=233$, as 
information of the appearance of a new gap structure at the elastic parametes 
$K_b$. This interpretation is wrong, because, if they calculated the model 
system with the same interaction strength for different particle numbers these 
artifacts will be vanished. The appearence of artifacts in the hull-function 
for the ground state defines the range of validity of the hull-function concept 
for several $\epsilon$ in the numerical calculation, which makes it necessary to
proof the hull-function structure for the same $\epsilon$ for several system
size. Relating to stationary state which is given by  Eq.(\ref{stationary}),
Fig. 5 shows the distance $L_0$ between the upper and the lower chain as a
function of the strength $\epsilon$ of the interaction force, which was
calculated for the particle numbers $N_a=233$ and $N_b=377$. The plot in 
Fig.\,5(a) shows the computation of $L_0$ for $\epsilon$ in range from $0.05$ to
$0.45$ and Fig.5(b) shows $L_0$ in the transition area. Note, the distance 
$L_0$ is remarkably insensitiv to the system size. For the system sizes
$\alpha=377/233$, $610/377$, and $1597/987$ we found a deviation for
$L_0=L_0(\alpha)$ smaller than $10^{-5}$. In the occurence of local artifacts
in the hull-functions for certain particle numbers and certain interaction
strength the deviation for $L_0 (\alpha)$ lies in the order $10^{-4}$.
Note, that in the ground state above the transition point $\epsilon_c$ also 
many local rearrangements take place which can be identified as first order
phase transitions.

\subsubsection*{{\normalsize {\bf 2. The order parameter}}}
\noindent
Now, we analyse the transition of breaking the analyticity near the threshold
$\epsilon_c$ in terms of second order phase transition. For this it is
convenient to introduce an order parameter, which is an indicator for the
occurence of a phase transition at a critical point. The order parameter 
distinguishs both phases which are involved in the phase transition, because
it is defined to be zero in one phase. Refering to 
the transition of breaking of analyticity the order parameter is defined in 
the discrete phase, which measures the discrete jumps in the hull-functions.
Therefore the order parameter is in the continuous phase for $\epsilon <
\epsilon_c$ equal to zero and in the discrete phase above $\epsilon_c$ unlike
zero. Here, we use the definition
\begin{equation}
\label{OP1}
OP_{m} = \frac{\displaystyle \int_{0}^{1} \left| \frac{\partial^m f(z)}{\partial z^m} 
\right|^2 dz}{\delta(0)} \, \, \, .
\end{equation}
Here, $\delta(0)$ behaves like 
$\delta(0) \sim \Delta z \, ^{-1}|_{\Delta z \rightarrow 0}$ . If the 
hull-function is smooth in the interval $[0,1]$, the integral becomes finite and
the order parameter vanished due to the prefactor $\delta(0) \, ^{-1}$. On the
other hand, if $f(z)$ has a discrete step at $z_0$, we obtain 
$\partial_z f(z) \sim \delta(z-z_0)$ and therefore
$\int_0^1 [ \partial_z f(z) ]^2 dz \sim \int_0^1 \delta (z-z_0) ^2 dz \sim
\delta(0) \int_0^1 \delta(z-z_0) \sim \delta(0)$, i.e. the order parameter
becomes finite. Due to the finite system size, which devides the region of 
integration in equidistant subintervals $\Delta z = 1/N_a$ ($\Delta z = 1/N_b$) 
for the upper (lower) chain, the integral in Eq.(\ref{OP1}) reduced to a sum 
over the particle numbers. Therefore, we have intrinsic discontinuities in the
numerical computation, because the hull-functions are given by a discrete set
of points at $\Delta z$ at all, which define the systematical error of the 
calculation. In order to smeared out the systematical intrinsic discontinuities, 
we choose for our numerical investigation $OP_4$ as order parameter. Hence, 
we get the following expression
\begin{equation}
\label{OP4}
OP \equiv OP_4 = \frac{\displaystyle \int_{0}^{1} \left| \frac{\partial^4 f(z)}{
\partial z^4} \right|^2 dz}{\delta (0)} \equiv \frac{1}{(\Delta z)^6}
\sum_{k=2}^{(N_b - 2) \atop N_a - 2} \big( \left| \, f_{k+2} - 4 \, f_{k+1}
 + 6 \, f_{k} - 4 f_{k-1} + f_{k-2} \, \right| \big)^2
\end{equation}
It should mentioned, that $OP$ shows necessary for all hull-functions 
$h_a$, $h_c$, $h_c$ and $h_d$ the same qualitativ behaviour. Therefore,
in order to show all basic features of the phase transition (transition of
breaking of analyticity) it is enough to calculate $OP$ (via Eq.(\ref{OP4}))  
only for the hull-function $h_a = h_a (z)$.

\subsubsection*{{\normalsize {\bf 3. The scaling behaviour}}}
\noindent
Next we have calculated the order parameter given by Eq.(\ref{OP4}) 
for different system sizes $\alpha=89/55$, $233/377$, $610/377$ and
$1597/987$. Figure 6(a) shows the order parameter as function of the interaction
strength $\epsilon$ in the transition region. We clearly see that the order
parameter $OP$ is changed from zero below a certain critical threshold $\epsilon_c$ 
to a value unlike zero above $\epsilon_c$ for all particle number ratios. This
is a certain hint that the transition of breaking of analyticity is a second
order phase transition. Using a finite-size scaling \cite{b23,b24} for the data
in Fig. 6(a) given by the scaling function
\begin{equation}
\label{scale_a}
OP (\epsilon - \epsilon_c, L) = L^{\frac{\beta}{\delta}} 
OP \left( L^{\frac{1}{\delta}} (\epsilon - \epsilon_c) \right)
\end{equation}
we find that the order parameter $OP$ exhibits a scaling behaviour in the
transition region shown in Fig. 6(b). That means, all curves of Fig. 6(a)
could mastered on one resulting curve (Fig. 6(b)). The system size $L$ is 
determined by the particle numbers $55$, $233$, $377$ and $987$.
In the transition region, which can be characterized by the reduced 
interaction strength $t=(\epsilon - \epsilon_c) / \epsilon_c$, we find that 
the data of Fig. 6(b) are consistent with the following power law for
$\epsilon > \epsilon_c$
\begin{equation}
\label{scale_b}
OP ( \epsilon - \epsilon_c) \sim L^{\frac{\beta}{\delta}} \left( 
\epsilon - \epsilon_c \right)^{\beta}
\end{equation}
with the critical exponents $\beta=0.239 \pm 0.05$, $\delta=0.687 \pm 0.05$ 
defining the ratio $\mu= \beta / \delta \approx 0.347 \pm 0.05$.  Due to the
finite-size scaling we obtain for the threshold interaction strength of 
the transition of breaking of analyticity the critical value 
$\epsilon_c=0.25575 \pm 0.001$.

\subsection*{{\normalsize {\bf B. The static friction}}}
\noindent
In the following we want to calculate the maximum static friction force $F_c$ of
the extended two-chain model refering to a pinning structure which is given in 
the incommensurate case for $\epsilon > \epsilon_c$. It is clear that in the 
presence of an applied forces $F^{ex} > 0$  ground states are no longer 
defined. From the physical intuition, we hope that one can adiabatically evolve
the underlying model system from the ground state by increasing $F^{ex}$ from
zero, because the threshold force $F_c$ is the largest depinning force $F^{ex}$
above no a longer stationary state exist, i.e., Eq.(\ref{stationary}) is no
longer valid. Like Ref. \cite{b10} the static friction is given by
\begin{equation}
\label{staticfriction}
F_R = - \sum_{i=1}^{N_a} \sum_{j=1}^{N_b} \left< \, F_{int} \left( 
x_i^a - x_j^b \right) \, \right>_t \equiv N_a \left< \, F^{ex} \, 
\right>_t 
\end{equation}
which can be obtained from the equation of motion (\ref{xakette})-(\ref{ybkette})
by summing over $i$ and averaging with respect to $t$. It should remarked,
that Eq.(\ref{staticfriction}) is also valid for the calculation of the
kinetic friction in a sliding state, i.e., for $F^{ex} > F_c$. For the
determination of the threshold force $F_c$, we choose especially the interaction
force $\epsilon=0.33$ and the system size $N_a=233$ and $N_b=377$. Note, the 
numerical calculations of $F_c$ for the extended two-chain model are very hard
and the computation time is very large. In a first calculation, we find that 
the maximum static friction force has approximately the value $F_c \approx 0.4$
but the numerical deviation is very large. Therefore, it takes much more efforts
in order to calculate the accurate value of the threshold force.


\section*{{\normalsize {\bf V. CONCLUSIONS }}}
\noindent
We have investigated the ground state for a two-dimensional extended two-chain
model with an incommensurate lattice structure. We have found that the 
structure of the hull-functions depends strongly on the strength $\epsilon$ of 
the interaction between the upper and the lower body. It was numerically 
observed the breaking of analyticity in the ground state for the critical 
value $\epsilon_c=0.25575$. Furthermore, it was shown that this type of phase
transition could be uniform described by an order parameter. This order
parameter representation allows a systematic analysis of the critical 
behaviour for $\epsilon > \epsilon_c$ due to the calculation of the critical
exponents. Consequently, the peculiar breaking of analyticity shows all features
of a second order phase transition.

A second interesting result of the ground state structure occurs in 
associtation with the numerical computation for different interaction strengths
as function of the rational system sizes. It has been shown that artifacts in
the hull-functions structure appear for certain rational particle numbers of a
fixed interaction strength $\epsilon$. This fact pointed out that the correct
ground state structure depends strongly on the choose of sufficient rational 
ratio of the particle numbers as a irrational number approximation for a 
certain range of the interaction force. 

Concerning the static properties of the extended two-chain model in the 
presence of an applied forces $F^{ex} > 0$ we have calculated the maximum 
static friction force $F_c$.

Finally, it should remarked, that the extended two-chain model is a simple model
structure , which established the macroscopic dry friction situation on the
microscopic level between atomically flat surfaces by using a simple driven 
mechanical many-body system. It is convenient to define the model as simple 
as possible in order to dicuss some basic features like the ground state
structure or the maximum static friction behaviour. Therefore we took in 
our investigation a simple relaxation dynamic into account and neglected 
thermal fluctuation ($T=0 \, K$). Surely, such kind of microscopic models 
could be always extended by additional thermal fluctuations representing by 
stochastical forces which leads to creep in the presence of applied forces.
Note, the vality of the underlying model system is given by the assumption of
simple harmonic particle interaction. It is clear that this assumption in many
real systems where large distortions occur. 

Note, there are one possible and very interesting extention of the extended
two-chain model by the introduction of quenched randomess which corresponds 
surface impurities. It is an open question, whether randomess destroy the
frictionless state in the incommensurate case for $\epsilon < \epsilon_c$
in the extended two-chain model like in the FK-model \cite{b7,b8}. Well, it is
an interesting question, whether randomess, which exists in any real system, 
leads in general to a finite maximum static friction force for any interaction
strength $\epsilon$. Refering to the compare whith experiments, another natural
extention of the present model is to proceed to a three-dimensional system, 
which is able to describe wearless dry friction between two atomically flat 
plans pinning in each case on a substrat and embeding in a three-dimensional 
space.    
    
In later works we want to investigate the threshold behaviour near $F_c$
especially for $F^{ex} > F_c$. It should be proofed whether the 
velocity-force characteristics shows a critical behaviour or not.
Well, for a complete understanding of the behaviour near the threshold $F_c$ 
a renormalization group calculation is needed, but surely very complicated for 
a deterministic mechanical system.

Note in the present paper, we have considered a simple overdamped motion.
Hence, it is also interesting to discuss the influence of inertia terms 
(underdamped dynamic) on the behaviour near the threshold $F_c$ and on the 
sliding state.

\section*{{\normalsize {\bf ACKNOWLEDGMENTS }}}
\noindent
This work has been supported by the Graduiertenkolleg 
"Defektstrukturbestimmte physikalische Eigenschaften"
and the Deutsche Forschungsgemeinschaft DFG (SFB 418).


\vspace*{9ex}


\begin{center}
{\bf FIGURES}
\end{center}

\vspace*{2ex}
\noindent
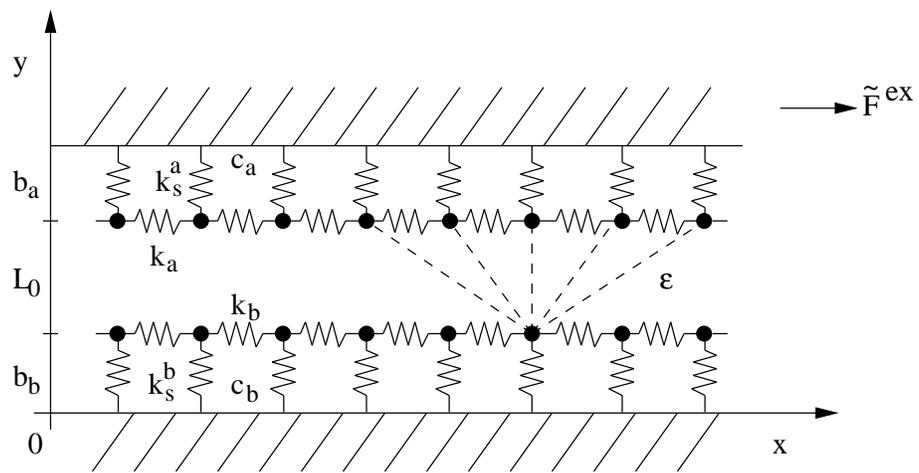
FIG. 1. A schematic picture of the two-dimensional extended two-chain model. 
The external force $\tilde{F}^{ex}$ is given by  the applied force $F^{Fex}$
which acts on each particle of the upper chain times the particle number $N_a$ 
of the upper chain where $N_a$, ($N_b$) is the particle number of upper (lower)
chain. \\

\vspace*{2ex}
\noindent
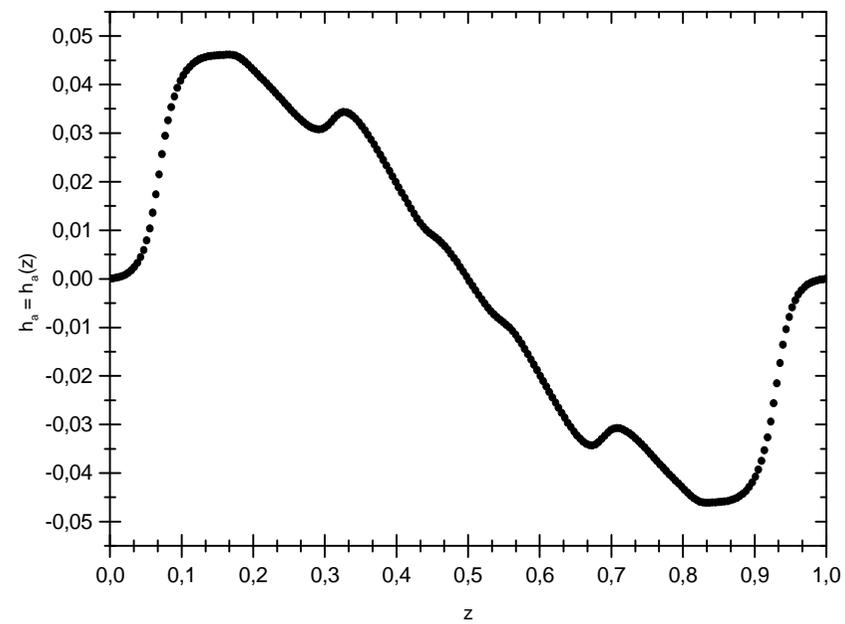
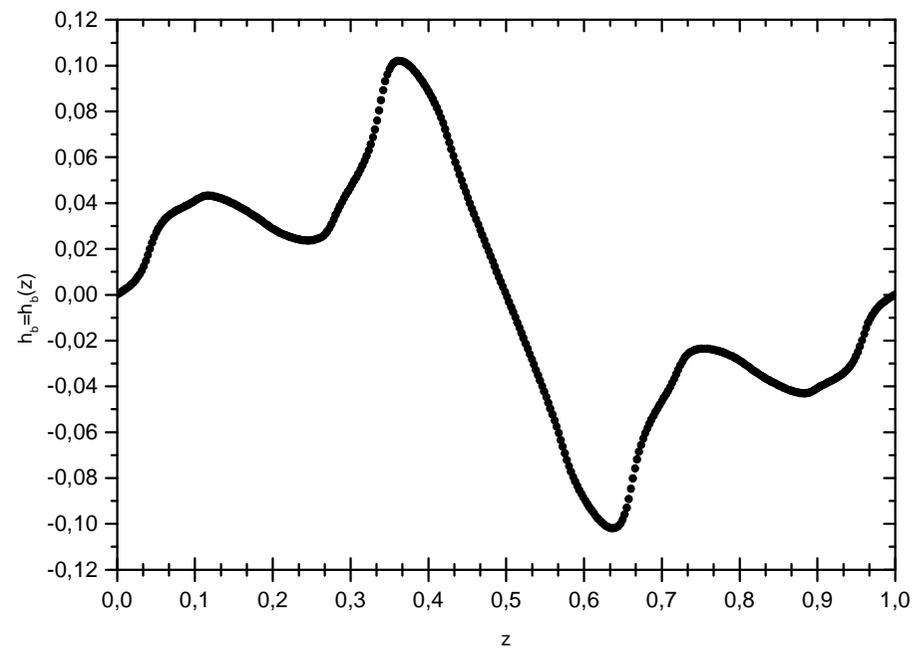
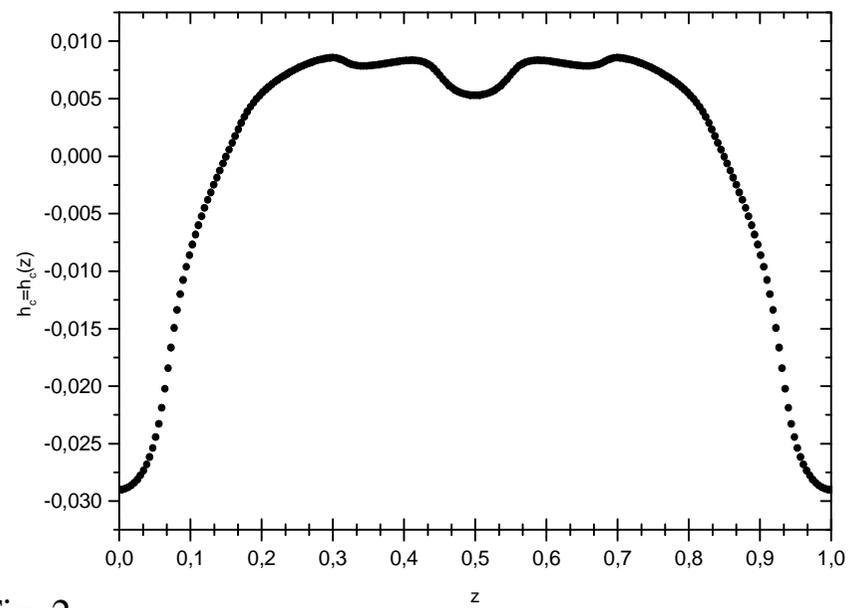
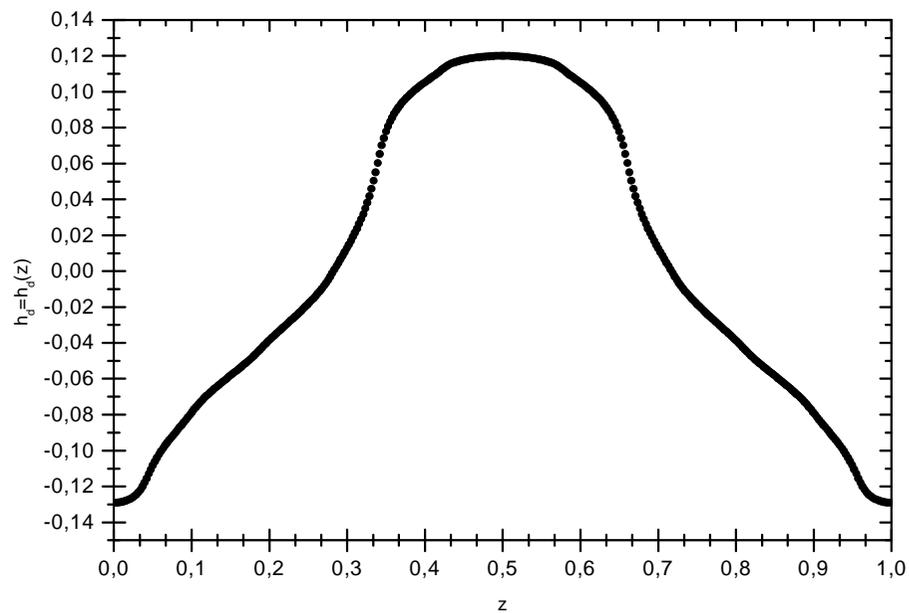
FIG. 2. The hull-functions of the ground state plotted for the particle 
numbers $N_a=233$ and $N_b=377$ with the interaction strength $\epsilon=0.22$.
For an uniform representation all hull-functions are renormalized by the mean
lattice spacing $c_b$ and $c_a$, respectively. Therefore, the substitution for
the hull-function $h_a$ and $h_c$ of the (a)-chain is 
$z \rightarrow z ' = z/c_b$, whereas for the hull-functions $h_b$ and $h_d$ of 
the (b)-chain it is $z \rightarrow z ' = z/c_a$. \\

\vspace*{2ex}
\noindent
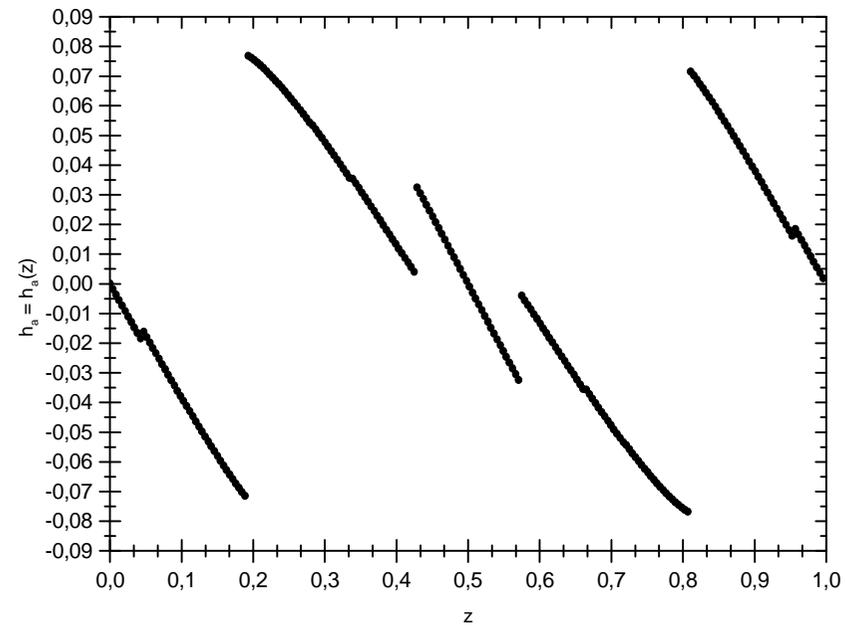
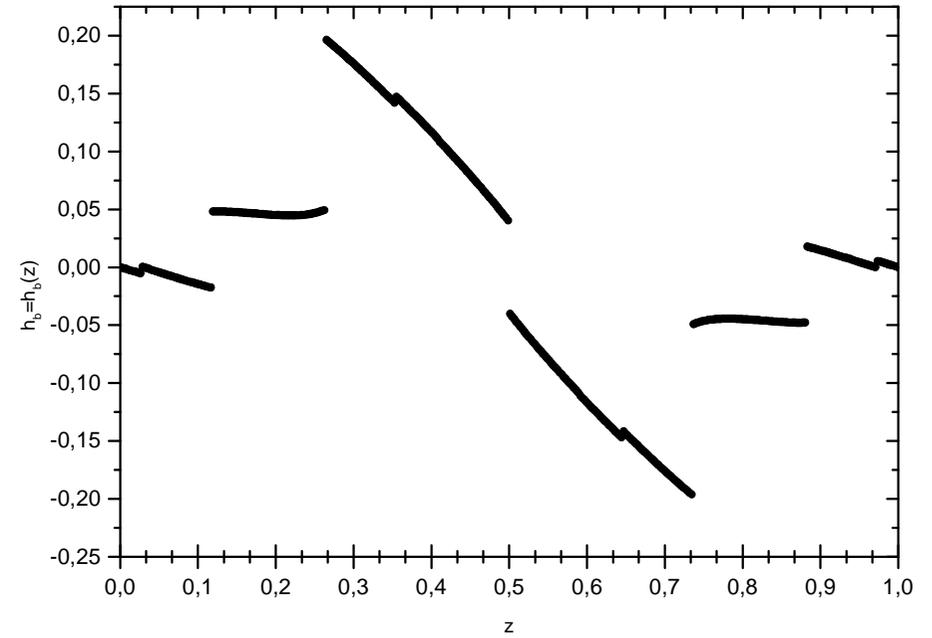
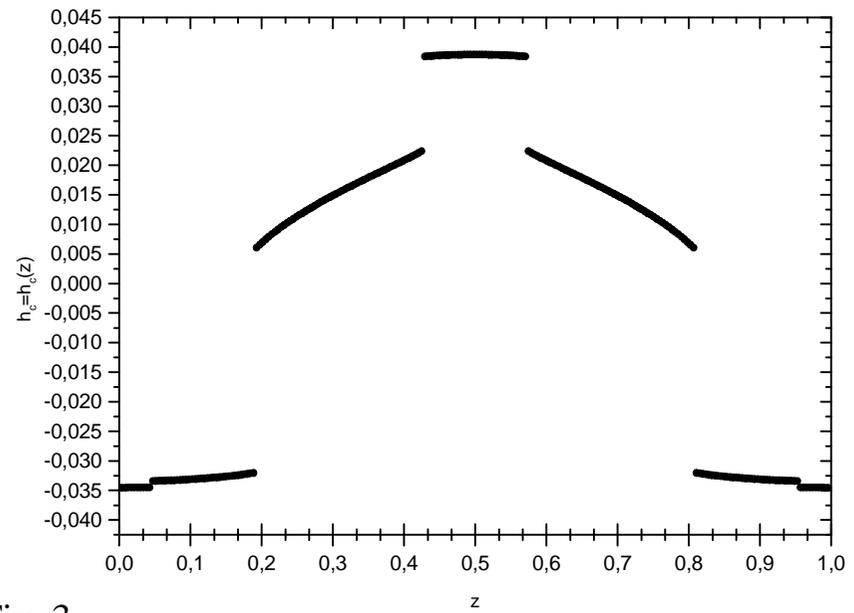
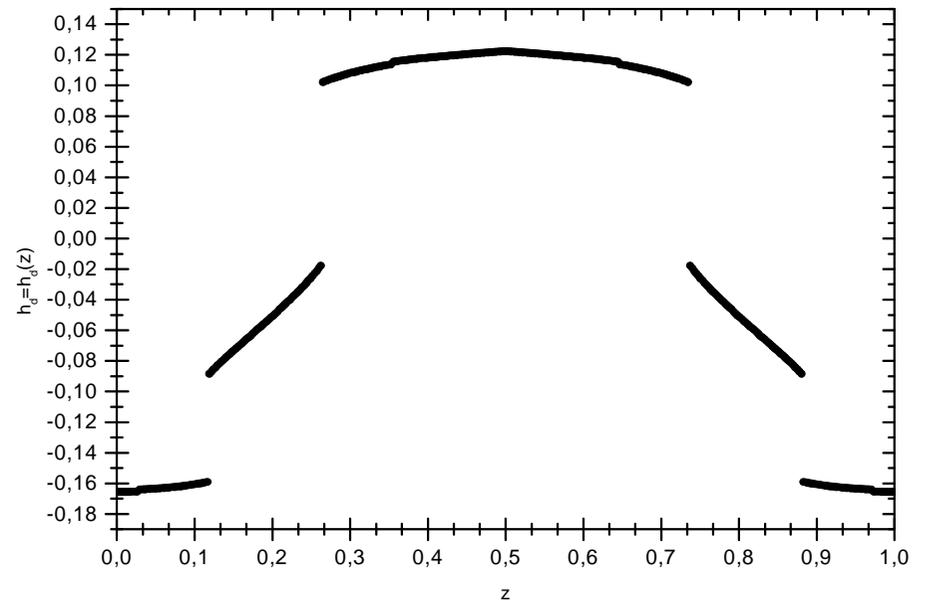
FIG. 3. Hull-functions of the ground state for $\epsilon=0.33$ and the
particle numbers $N_a=233$ and $N_b=377$. \\

\vspace*{2ex}
\noindent
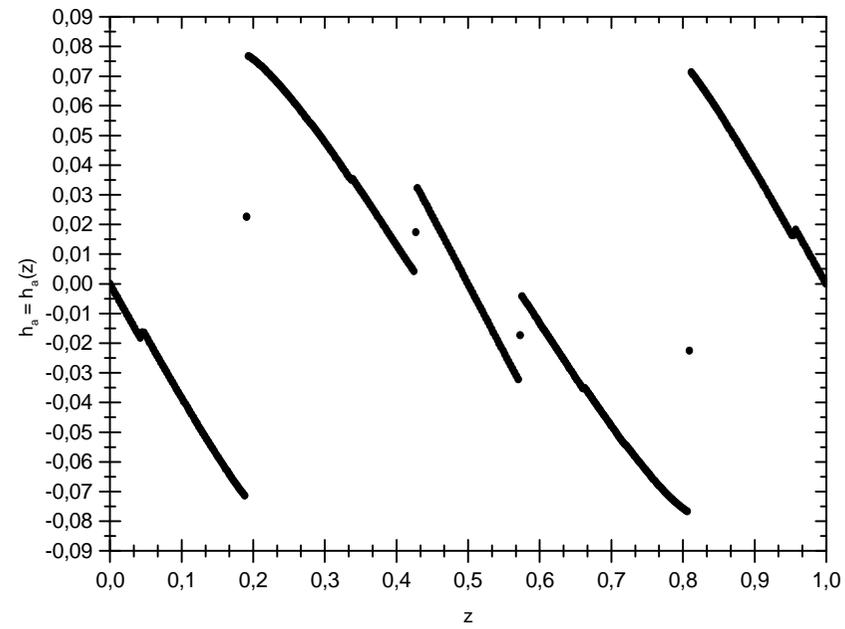
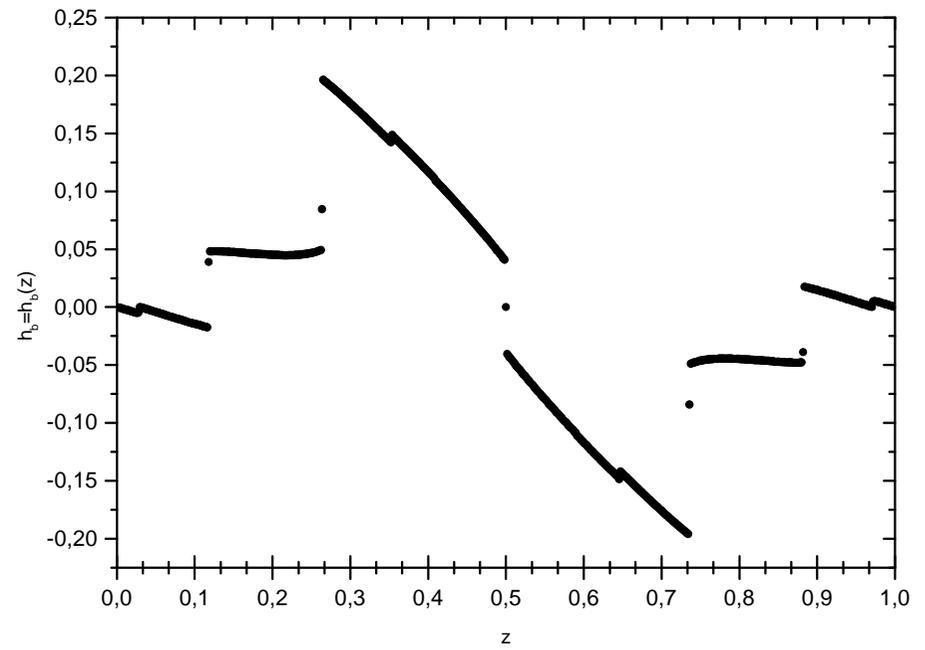
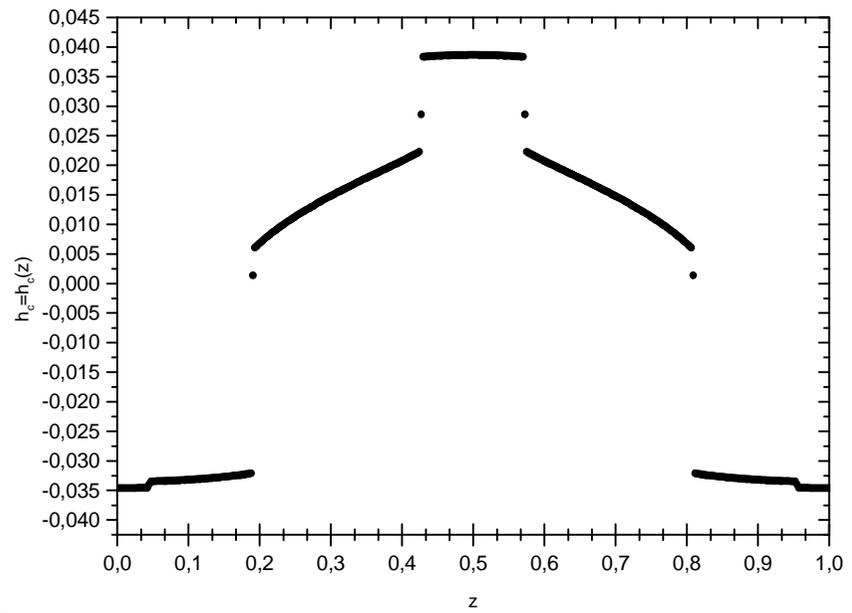
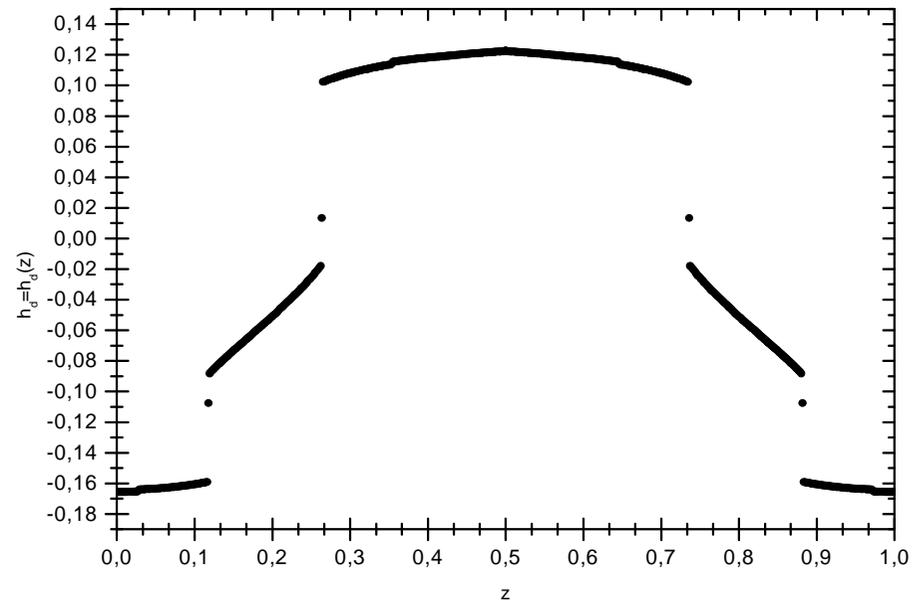
FIG. 4. Artifacts in the hull-functions for a bad choice of the particle
numbers, e.g., given by $N_a=377$ and $N_b=610$ and the interaction strength
$\epsilon=0.33$. \\

\vspace*{2ex}
\noindent
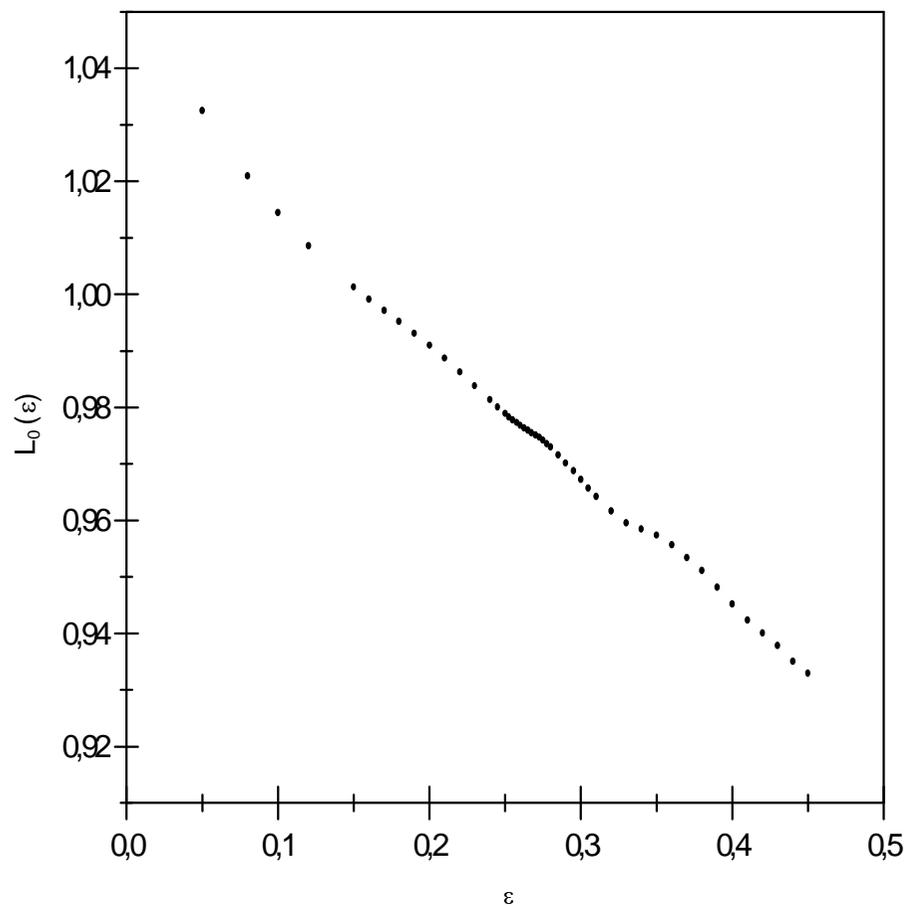
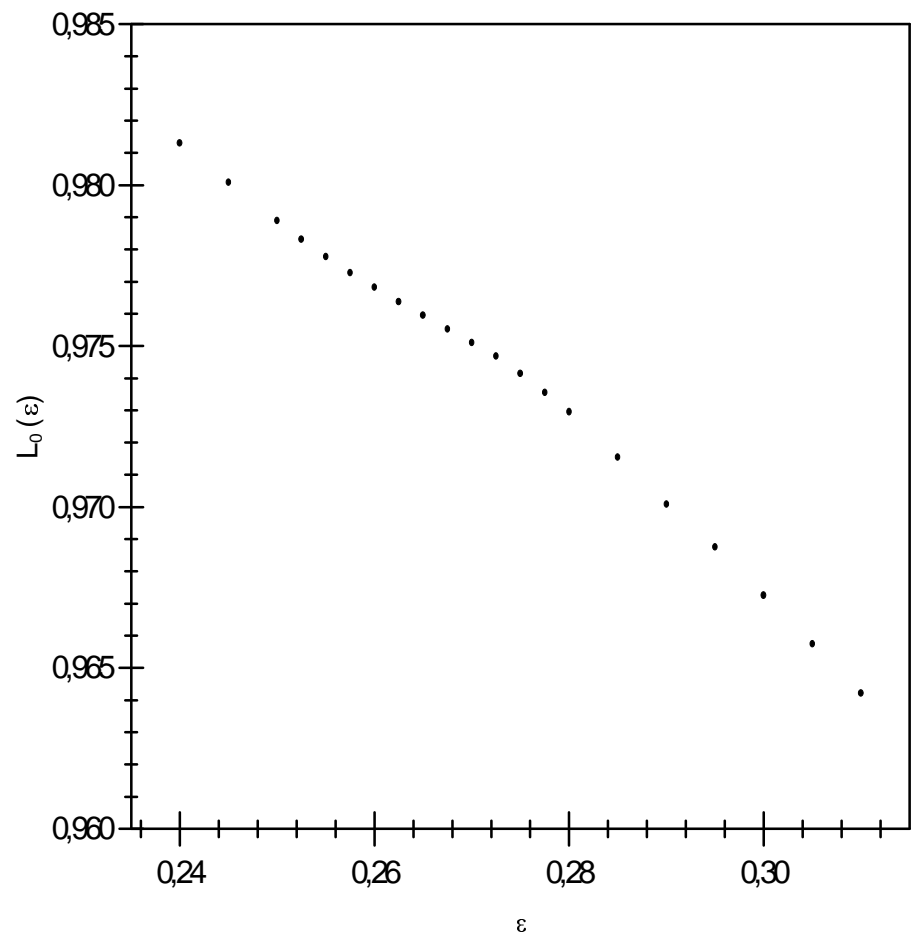
FIG. 5. The interaction distance $L_0$ in $y$-direction between the upper and 
lower chain as function of the interaction strength $\epsilon$. The distance 
$L_0$ is plotted in different ranges (a)  from $\epsilon=0.05, \cdots ,0.4$ 
and (b) from $\epsilon=0.24, \cdots ,0.31$ (transition range).   \\

\vspace*{2ex}
\noindent
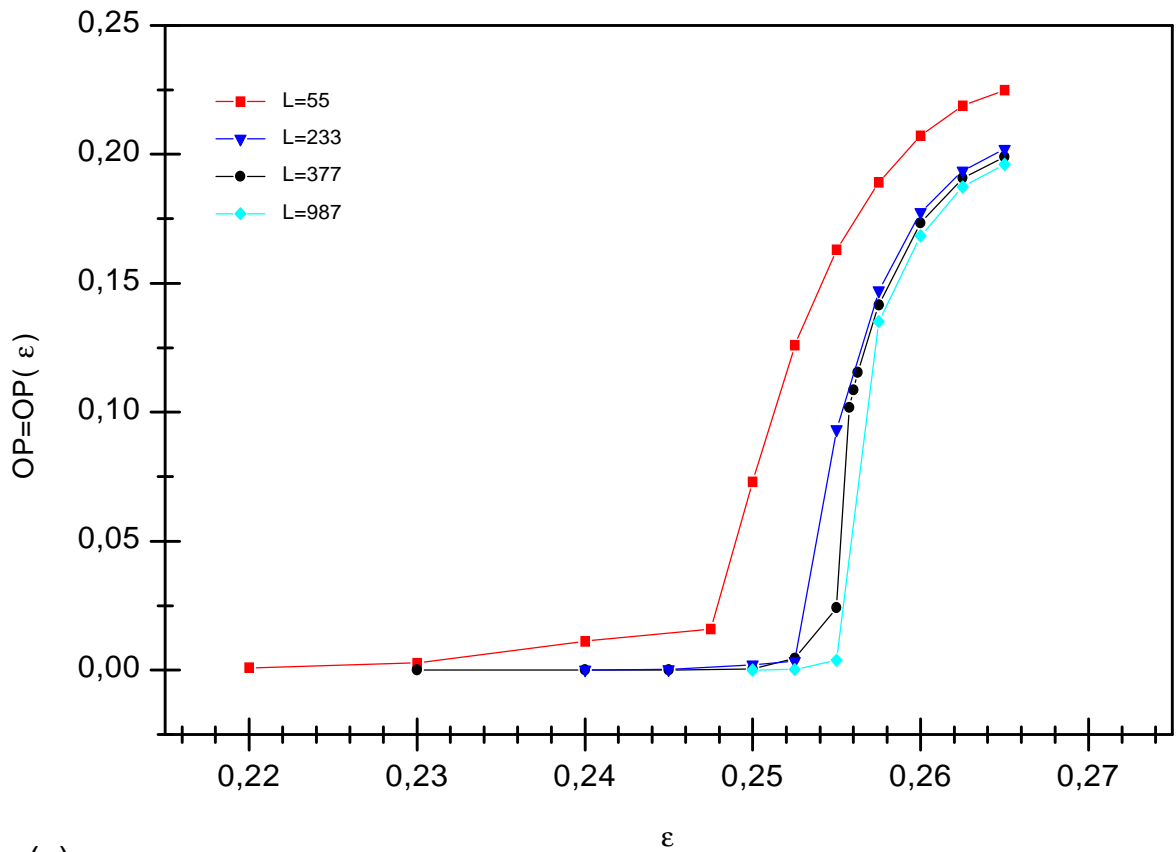
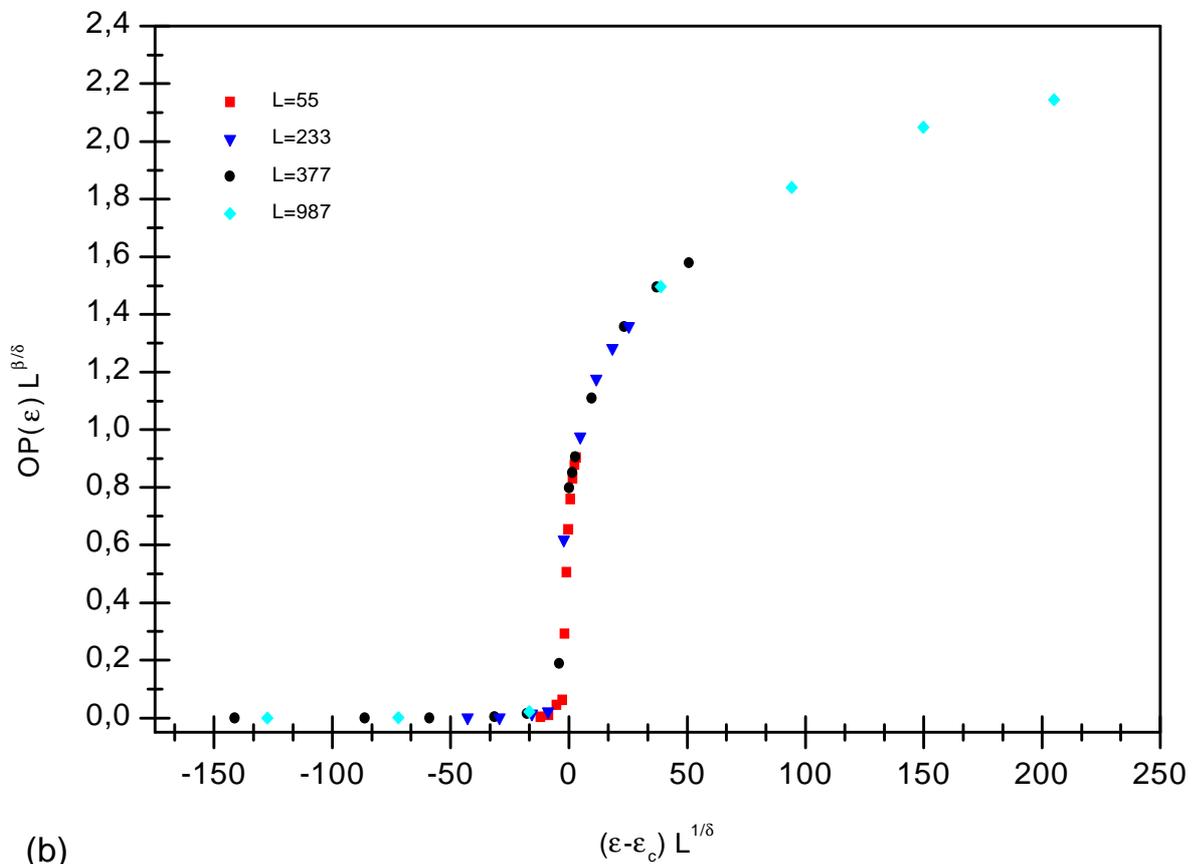
FIG. 6. In the computation it is convenient to rescale the order parameter by 
$OP$ $\rightarrow$ $OP \cdot (\Delta z)^6$. The order parameter 
$OP=OP(\epsilon - \epsilon_c)$ describes the critical behaviour near the 
threshold $\epsilon_c$. The graph (a) shows the
behaviour of the order parameter for different particel numbers and (b) is the
finite-size scaling plot of the mastered order parameter, where 
$OP(\epsilon) L^{\frac{\beta}{\delta}}$ is plotted versus 
$(\epsilon - \epsilon_c) L^{\frac{1}{\delta}}$.  \\

\end{document}